\documentclass[conference]{IEEEtran}
\IEEEoverridecommandlockouts
\usepackage{cite}
\usepackage{amsmath,amssymb,amsfonts}
\usepackage{algorithmic}
\usepackage{graphicx}
\usepackage{textcomp}
\usepackage{makecell}

\makeatletter
\let\@ORGmakecaption\@makecaption \long\def\@makecaption#1#2{\@ORGmakecaption{#1}{#2}\vskip\belowcaptionskip\relax} \makeatother

\let\oldequation\equation
\let\oldendequation\endequation
\renewenvironment{equation}
{\linenomathNonumbers\oldequation}
{\oldendequation\endlinenomath}

\input epsf
\usepackage{multirow}
\usepackage{balance} 
\usepackage{stfloats}
\usepackage{xcolor}
\usepackage{listings}
\usepackage[numbers,sort&compress]{natbib}
\usepackage[switch]{lineno}
\usepackage{indentfirst}
\usepackage{url}
\newtheorem{basis}{\textbf{Annotation Basis}}
\newtheorem{combination}{\textbf{Combination Rule}}

\definecolor{dkgreen}{rgb}{0,0.6,0}
\definecolor{gray}{rgb}{0.5,0.5,0.5}
\definecolor{mauve}{rgb}{0.58,0,0.82}

\newcommand{\ToolName}[0]{\texttt{JustinANN}}

\def\BibTeX{{\rm B\kern-.05em{\sc i\kern-.025em b}\kern-.08em
    T\kern-.1667em\lower.7ex\hbox{E}\kern-.125emX}}
    
\begin{document}

\title{\ToolName{}: Realistic Test Generation for Java Programs Driven by Annotations}

\author{
Baoquan Cui, Rong Qu, Jian Zhang \\
State Key Laboratory of Computer Science, Key Laboratory of System Software \\Institute of Software, Chinese Academy of Sciences, Beijing, China
\\Email: \{cuibq, qurong, zj\}@ios.ac.cn



}


\maketitle
\begin{abstract}
Automated test case generation is important.
However, the automatically generated test input does not always make sense, and the automated assertion is difficult to validate against the program under test.
In this paper, we propose \ToolName{}, a flexible and scalable tool to generate test cases for Java programs, providing realistic test inputs and assertions.
We have observed that, in practice, Java programs contain a large number of annotations from programs, which can be considered as part of the user specification.
We design a systematic annotation set with 7 kinds of annotations and 4 combination rules based on them to modify complex Java objects.
Annotations that modify the fields or return variables of methods can be used to generate assertions that represent the true intent of the program, and the ones that modify the input parameters can be used to generate test inputs that match the real business requirement.
We have conducted experiments to evaluate the approach on open source Java programs.
The results show that the annotations and their combinations designed in this paper are compatible with existing annotations; our approach is easier to generate test data in, on and outside the boundaries of the requirement domain; and it also helps to find program defects.

\end{abstract}
\IEEEoverridecommandlockouts
\vspace{1.5ex}

%
\IEEEpeerreviewmaketitle

\lstset{numbers=left,
    language=Java,
	basicstyle = \ttfamily\footnotesize, 
	numberstyle=\tiny, 
	columns=flexible,
	keywordstyle=\color{blue}\bf, 
    morekeywords={@interface},
    commentstyle=\color{dkgreen},
	frame=lines,
	breaklines, 
	xleftmargin=2em,
	extendedchars=false, 
	tabsize=3,
}

\section{Introduction}

Unit testing is an important software testing method to ensure software quality  as it helps to detect bugs and prevent errors/exceptions.
Automatic test case generation for Java programs is increasingly being adopted by developers because it is powerful and convenient.

Existing approaches typically use randomness ~\cite{Randoop} or search-based software testing (SBST)  approach ~\cite{EvoSuite} when generating test cases.
Random testing is a black-box software testing technique where programs are tested by generating random inputs.
SBST is the application of optimizing search techniques (for example, genetic algorithms with seeds and mutation) to solve testing problems.

However, the cases they generate typically suffer from two problems.
Firstly, the input data generated by existing approaches does not always match the business scenario, making it difficult to understand and maintain the test cases.
For example, consider a data generation case that needs to generate an integer parameter representing age in the real world.
The randomness approach in \texttt{Randoop}~\cite{Randoop} will select a random value between the minimum and maximum integer values. 
Assuming that the reasonable range of human ages is 0-150, then the probability of randomly generating a reasonable age is $151/2^{32}$, which is almost impossible unless repeated as many times as necessary, i.e., it is difficult for the random approach to generate a reasonable age.
The SBST approach, as used in \texttt{EvoSuite}~\cite{EvoSuite}, 
is also difficult to achieve a reasonable age unless a good enough seed and age mutation process can be guaranteed.

The other problem is about assertion generation. 
Existing tools first execute the test case without assertion to get the execution result, and then use the execution result to generate assertion statements to obtain the test case with assertion.
This approach may be applicable to regression testing, but the assertion does not validate the intent of the program function as it is ``self-judgment", which is far from the actual assertion of the return variable (\textit{r}).
For example, if the return value of the test method is 4, at this time they will automatically add the assertion 
$$assert(r == 4 )$$
at the end of the test case for the given inputs, which cannot actually verify that the return value 4 is expected or not. 
That is to say, they cannot generate reasonable assertions.

In fact, annotations are widely used in Java programs in practice~\cite{Annotations,Annotations2} and  we have observed that they can be used to aid in test case generation.
An annotation is a special interface/class used to modify various elements in a class, such as a class, field, method, parameter, etc.
It can be compiled as a part of the program and accessed statically via the bytecode or even dynamically at run-time.
They can modify the elements in the class with the real-world business information which may indicate the range of an integer parameter, or indicate that an input string matches an email address format.

Although the annotation is helpful for test case generation, there are also some challenges.
The first one is how to be compatible with existing annotations as there are so many annotations already.
The second is how it can be easily used to meet the requirements for complex object modification.

In this paper, we propose an annotation-based approach to guide test case generation with the realistic input data  and the assertion, and have developed a flexible and scalable tool named \ToolName{} based on it.
In order to accurately express the basic characteristics of variables, we have designed 7 basic annotations to cover the existing common annotations, and  proposed 4 combination rules to improve their expressiveness for complex objects.
Annotations modifying method parameters are used to generate test cases; annotations modifying fields and method return variables are used to generate assertion statements.

In summary, the contributions of this paper are as follows: 
\begin{itemize}
\item \textbf{Approach.} We propose an approach to generate realistic test case based on annotations, design 7 kinds of annotation bases to describe the basic feature for the variable of the basic type and provide 4 combination rules based on them to extend their expressive power.
\item \textbf{Tool.} We have developed a tool named \ToolName{} to implement our approach and evaluate it on real world Java programs. 
Experiments show that many existing annotations are compatible with the annotations designed by us and their combinations; it is easier to generate input data in, on and out of the required range; and it also helps to find program defects.
\end{itemize}

\section*{Acknowledgement}
\ToolName{} is publicly available anonymously  \footnote{https://anonymous.4open.science/r/JustinANN-BB53}. 
Users can use it  via both the command line and in plug-in mode in the Eclipse IDE.

\section{Background}
In this section, we will introduce the annotation in Java and its usage. And an example will show the limitations of the existing test case generation approaches and how the annotation will help to generate test cases with the realistic input and assertion.
\subsection{Annotation}
Annotation is a special interface marked by the symbol `@' in front of the class name when defined in Java, used to modify the syntax metadata of elements in the class file. 
An annotation type declaration is similar to a normal interface declaration (Listing~\ref{lstlisting:An Example of Annotation}, \texttt{Range}). 
It can have no method declarations and method declarations must not have any parameters.
But the return type of its method is restricted to primitives, String, Class, Enum, annotations, and arrays of the preceding types.
We also call its methods attributes, because the instantiation of method return values is similar to an assignment rather than a typical functional implementation (``\texttt{@Range(min=0, max=150)}", Listing~\ref{lstlisting:An Example of Annotation}, line 12).
Since version 1.5, the JDK has provided 4 meta-annotations to modify annotations, i.e., $@Retention, @Target, @Documented$ and $@Inherited$. 
$@Retention$ indicates the cycle of the annotation, i.e. whether it is in the Java file or in the class file, or whether it can be retrieved at runtime. 
For example, in Listing~\ref{lstlisting:An Example of Annotation}, the $@Retention$ (line 2) shows that the annotation \textit{Range} modified by it can be complied into the bytecode class when it is used (lines 11 and 12). 
$@Target$ marks which elements in the class file can be annotated with this annotation, such as a class, method, parameter, or field. 
For instance, the annotation \textit{Range} in Listing~\ref{lstlisting:An Example of Annotation} can be used to modify a parameter (line 12) as itself is modified by the meta-annotation $@Target$ with a value \textit{PARAMETER} and a method (line 11) as with a value \textit{METHOD}.
$@Documented$ declares whether it can be recorded by the Java Document, and $@Inherited$ represents that the annotation can be inherited by other annotations. 

\textbf{Annotation Usage}.
Annotations do not directly affect the semantics of programs, but they do affect the way programs are handled by tools and libraries, which in turn can affect the semantics of the running program. 
Annotations can be read from source files, from class files or at runtime, respectively.
A systematic investigation of 1,094 notable open source Java programs has shown that nearly three quarters (73.71\%) of the assigned annotation values are typed as string, and in some cases the string content could actually be better typed (for example, as class or primitive) ~\cite{Annotations}.
Another work on a corpus of 106 open source Java systems with an annotation count of over 160,000 has found that developers use both predefined annotations and annotations defined by external frameworks, mostly annotations dedicated to persistence and testing ~\cite{Annotations2}.
For example, 
JavaBeans Validation, provided by JavaEE, contains annotations such as \textit{@Max, @Min, @Size} and so on for validating constraints placed on fields, methods or classes, based on the Java Specification Requests (JSR), such as JSR-303~\cite{JSR-303}.
It has been used by 6,623 artifacts in the Maven repository \textit{MvnRepository} ~\cite{Artifacts-using-Bean-Validation-API}, including the popular frameworks, Spring Context, Spring Web, and so on.
For these reasons, the annotation is widely used by developers in Java programs.
Next, we will show how it helps us to generate test cases.

\subsection{Motivation Example}
\label{subsection: Motivation Example}
\begin{figure}[t]
    \centering

\begin{lstlisting}[
caption={An Example of Annotation},
label={lstlisting:An Example of Annotation}]
@Target({ METHOD, PARAMETER }) //meta-annotation
@Retention(RUNTIME) //meta-annotation
public @interface Range{//Annotation Declaration
    int min() default 0;
    int max() default 10000;
}

public class Ticket{
    int value;
    // Usage of the annotation Range
    @Range(min=80, max=160)
    int discount(@Range(min=0, max=150) int age){ 
        if(age <= 12 || age >= 60){
            return value * 0.5; 
        }
        return value;
    }
}
\end{lstlisting} 
\end{figure}

Listing~\ref{lstlisting:An Example of Annotation} shows an example with ticket discount function (lines 11-17): seniors (over 60 years old) and children (under 12 years old) can enjoy half price discount.
For the random approach, it may produce an arbitrary integer value  between the upper and lower bounds of the range expressed by the Java integer type.
The value  is hardly a reasonable age and may even be  a negative value (i.e, -3846), which is unacceptable to test suites and unreadable to  test case maintainers.
Furthermore, for the SBST approach, it may obtain a seed from the predicate about the  \textit{age} parameter, $$age \leq 12 \lor age \geq 60$$ (line 13), then it produces a solution via a solver such that the predicate is satisfied.
This value may be far away from a normal human age (i.e, 500) and may be even negative too, which also makes it difficult to find a solution for a reasonable age.

If we take the annotation \textit{@Range} modifying the parameter \textit{age} into account (Listing~\ref{lstlisting:An Example of Annotation} , line 12), as the user specification, the range (0-150) of human age will be easily obtained, then combined the result from SBST, we can get a candidate set of the boundary values, including with specification values and  SBST solutions, such as 
$$\{0, 12, 60,150\}.$$
The input candidates can be generated from the bounds themselves and the range between adjacent bounds, such as 
$$\{-12 , 0, 8, 12, 34, 60, 78, 150, 500\}.$$
Obviously, these candidates contain many reasonable age values and also exceptional ones, which is very friendly to test case readability and maintainability for users.

Moreover, for the existing test case generation tools, they generate assertions for the return value \textit{r} by executing the case without any assertion at first, obtaining a result value (assuming 50), adding an assertion of equality between the return variable and the result value, 
$$ r = discount(...);  \qquad assert(r == 50),$$ 
and finally outputing a test case with the assertion.
However, it is not known whether this assertion is the expected behaviour of the method \textit{discount(...)}.
If we take the annotation \textit{@Range} modifying the method \textit{discount(...)} into consideration (Listing~\ref{lstlisting:An Example of Annotation} , line 11), then we can generate an assertion, 
$$r = discount(...);  \qquad assert(r \geq 80 \ \&\& \  r \leq 160).$$
At this point, if a test case returns a value of 50, then the assertion will tell the developer that this is a buggy method.

\section{Technique}
\begin{figure*}[t]
\centering
\includegraphics[scale=0.5]{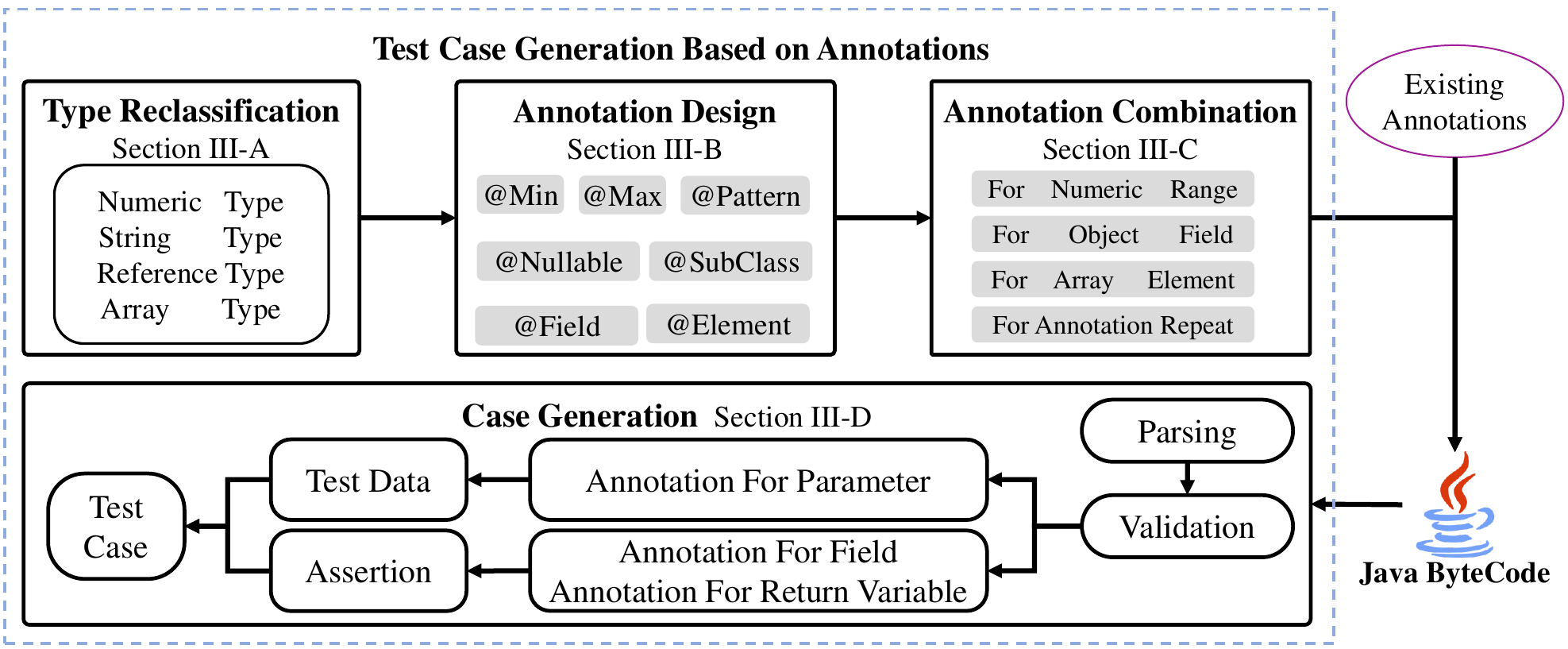}
\caption{Overview of Our Approach}
\label{Figure: Overview}
\end{figure*}

As mentioned before, the annotation is widely used in the Java program while there are two challenges to use them to help test case generation:  how to be compatible with existing annotations and how to modify complex objects.

To counter the challenges, we collect statistics on the semantics of annotations in real projects to observe the annotations.
And we have found that the annotations which can be used to help for test case generation are strongly related to the variable type, whether they come from different platforms  or use different names.
For example, an annotation named \texttt{Min} always modifies an integer variable, which may be an integer variable or an integer field of a complex object, whether the annotation \texttt{Min} comes from different platforms, such as \textit{javax.*} and \textit{com.google.*}.
For another example, two annotations with different names, \textit{Null} or \textit{Nil}, both express an object may be a \textit{NULL} value.
That is, we can start from the variable type, design basic annotations to express common semantics, and then interpret existing annotations into corresponding semantics (compatible and scalable).
With the basic annotations, we can use combination to increase their expressive power, i.e. flexibly explain the modification of complex objects by combination.

\ToolName{} is designed as a flexible and scalable tool to generate test cases with realistic inputs and assertions.
Figure ~\ref{Figure: Overview} shows an overview of our approach, where the annotations are designed based on the Java types.
Although the Java language traditionally has two categories of variable types, primitive and reference, we have reclassified the types of the Java language in order to modify program elements with fewer and more efficient annotations in practice (Section ~\ref{subsection: Type Reclassification}).
Then, for each classified type, we have designed the basic annotation, which we call annotation basis in this paper, to express its basic characteristics (Section ~\ref{subsection: Annotation Basis Design}).
Thirdly, in order to meet the needs of complex constraint expressions, we have combined some of these annotation bases with  combination rules (Section ~\ref{subsection: Annotation Combination}).
Finally, we generate realistic test cases based on the annotations designed and the combination rules (Section ~\ref{subsection: Implementation}).

\subsection{Type Reclassification}
\label{subsection: Type Reclassification}
Java is a typical example of a statically typed language, where it is a compile-time error to assign a value of an incompatible type to a variable.
There are traditionally two categories of variable types in Java: the primitive type and the non-primitive type.
There are 8 primitive data types in Java: boolean, char, byte, int, short, long, float, and double.
And the non-primitive types include the array type and the non-array type.
Such a complete classification is necessary and efficient for Java programs. 
However, there is redundancy in the guidance for constructing the annotation base. 
For example, it is not necessary to define different Max annotations (\textit{@MaxS, @MaxI}) for short and int types to modify variables of these types.
Therefore, we simplify the classification.

\textbf{Primitive type.}
If we restrict the value of the boolean type of a variable, this is equivalent to making the variable a constant, since it has only two values (TRUE and FALSE), and the semantics are lost at this point, so we ignore it.
As for the remaining seven types, although they represent different types, they can all be classified as numeric types (char and int types are convertible in Java).
\textbf{Non-primitive type.}
There is a significant difference between the array type and the non-array type in terms of value restrictions on variables.
The former requires more index and element specifications.
For convenience, we also refer to the non-array type as the reference type defined as a class or interface.
Therefore, we reclassify the type as follows.
$$
Type := NumericType \ | \ String \ | \ RefType \ | \ ArrayType
$$
$$
ArrayType := (NumericType \  | \ String \ | \ RefType) []
$$
For each different type, we can define a limited annotation basis to modify the variables via their types.

\subsection{Annotation Basis Design}
\label{subsection: Annotation Basis Design}
For each reclassified type, we have designed the corresponding annotation to express the most basic restriction of the variables defined by that type.

(1) For a numeric type, the most basic constraints are the upper and lower bounds, so we use annotations, \textit{@Min} and \textit{@Max}, to express them respectively.
(2) It is well known that the regular expression can characterise a string with rich semantics. 
So we designed the annotation \textit{@Pattern} to express the string constraint.
(3)For an object, its nullity is important since the \texttt{NULL} input may cause a fatal error and  checking \texttt{NULL} everywhere makes the programmer nauseous. 
An improvement is  to mark whether an input object is nullable.
\begin{basis}
\textbf{@Min.}
\textit{
The annotation \textit{@Min} is an indication of the minimum value of a numeric variable.
The annotated element must be a number whose value must be larger than or equal to the specified minimum.}
\end{basis}
\begin{basis} \textbf{@Max.}
\textit{The annotation \textit{@Max} is an indication of the maximum value of a numeric variable.
The annotated element must be a number whose value must be smaller than or equal to the specified maximum.}
\end{basis}
\begin{basis}
\textbf{@Pattern.} 
\textit{The annotation \textit{@Pattern} is an indication of the matching pattern of a String variable.
The annotated String variable must match the following regular expression.}
\end{basis}

\begin{basis}
\textbf{@Nullable.}
\textit{
The annotation @Nullable is an indication that the value of a reference variable may be \texttt{NULL}.
The annotated element may be a \texttt{NULL} or not.}
\end{basis}
Listing ~\ref{lstlisting:Declaration and Usage of the Annotation Basis} shows their declaration and usage, for example, modifying the numeric parameter \textit{i} with a minimum value of 5 and a parameter \textit{j} with a maximum value of 20 (line 7).

\begin{figure}[t]
    \centering
\begin{lstlisting}[
caption={Declaration and Usage of the Annotation Basis},
label={lstlisting:Declaration and Usage of the Annotation Basis}]
//Declaration
public @interface Min{ int value();}
public @interface Max{ int value();}
public @interface Pattern{ String value();}
public @interface Nullable{ }
//Usge
void foo(@Min(5) int i, @Max(20) int j, @Pattern("[0-9]*") String s, @Nullable Object o){...} 
\end{lstlisting}
\end{figure}

\subsection{Annotation Combination}
\label{subsection: Annotation Combination}

The above annotations are only for variables of basic types (String can also be regarded as a special type). 
For a complex object in the object-oriented language, we need to consider its  encapsulation and polymorphism  characteristics. 
The former one is about the encapsulation of the input data, i.e., the field of the input object.
We design the annotation \textit{@Field} to express the restriction.
The latter is about the subclass of the input type, and we design the annotation \textit{@SubClass} for it.
Among them, array is a special type of object. 
Its element is similar to the field of an object. 
To avoid confusion, we design a similar annotation \textit{@Element} to express the restriction to its element.
Moreover, when the annotations, \textit{@Max} and \textit{@Min}, are used to modify array variables, they can be interpreted as restrictions on the length of the array, and there will be no ambiguity.
Obviously, the annotations, \textit{@Field} and \textit{@Element}, of them can not be used in isolation.
Because they only indicate which sub-object will the restriction is for and must be combined with annotations for the basic types.
\begin{basis}
\textbf{@Field.}
\textit{The annotation @Field is an indication that the restriction focuses on the field of the variable, with an attribute representing the name of the field.}
\end{basis}
\begin{basis}
\textbf{@SubClass.} 
\textit{ The annotation @SubClass is an 
indication of the possible subclass of the class type of the variable, with an attribute representing the full class name (with its package name)  of its subclass.
}
\end{basis}
\begin{basis}
\textbf{@Element.}
\textit{The annotation @Element is an indication that the restriction focuses on the specific element of the array, with an attribute representing the index of the element.}
\end{basis}

Inspired by the way Java class types are constructed from primitive types and nested by reference types, we also combine the annotation bases to unleash their powerful expressive capabilities.
First of all, we can combine the annotation bases, \textit{@Min} and  \textit{@Max}, to form an interval, which is used to express the range of numerical variables. 
\begin{combination}
\textit{
A numeric range is a combination of the @Min and @Max annotation bases, representing the lower and upper bounds of a variable.
}
\end{combination}

In addition, expressing constraints on objects is one of the challenges of test generation.
As mentioned earlier, we already have the \textit{@Field} annotation to modify a member of the object, but there is no specific constraint on it. 
This problem can be solved by using the \textit{@Field} annotation in combination with other annotations. 
\begin{combination}
\textbf{}
\textit{
The \textit{@Field} annotation indicates the specific field for modification, and other annotations indicate specific constraints: the use of numeric and string types is as described in Section~\ref{subsection: Annotation Basis Design}; the modification of a reference type variable can be nested according to the above process (although this is not recommended due to its readability and maintainability).
}
\end{combination}
Moreover, the element restriction representation of an array is similar to the field representation of a class.
\begin{combination}
\textbf{}
\textit{
The \textit{@Element} annotation indicates the specific element in the array for modification, and other annotations indicate specific constraints as described in Section ~\ref{subsection: Annotation Combination}.
}
\end{combination}
Finally, a variable may need to be modified with the same annotation or annotation combination multiple times.
For example, a string variable for contact information may be an email address or a phone number.
\begin{combination}
\textit{
A list is a single annotation or repeated occurrences of the same combination, indicating multiple similar constraints on a variable.
The combination should be defined explicitly as an annotation container to clearly distinguish how the annotations are combined.
}
\end{combination}

\begin{figure}[t]
    \centering
\begin{lstlisting}[
caption={Example for the Combination of Annotations},
label={lstlisting: Example for the Combination of Annotations}]
//Annotation Declaration
public @interface Field{ String value(); } 
public @interface SubClass{ Class<?> value(); }
public @interface Element{int value();}
// class Declaration
class Person{ int age; ... }
class Student extends Person{ ... }
class Teacher extends Person{ ... }

//Combination 1: the input i is between 5 and 20
void foo(@Min(5) @Max(20) int i){...}
//Combination 2: the field age of  the input p is between 5 and 20
void foo(@Field("age") @Min(5) @Max(20) Person p){...}
//Combination 3: the field age of the second element in the input ps is between 5 and 20
void foo(@Element(2) @Field("age") @Min(5) @Max(20) Person[] ps){...}
//Combination 4: the string type parameter s may be a string of lowercase letters, or a string of uppercase letters or a string of numbers.
void foo(@Pattern("[0-9]*") @Pattern("[a-z]*") @Pattern("[A-Z]*") String s){...}

//Powerful Combination: the second element of the input array ps has type Student and its field age is between 5 and 20 
void foo(@Element(2) @SubClass(Student.class) @Field("age") @Min(5) @Max(22) Person[] ps){
...}

\end{lstlisting}
\end{figure}

\textbf{Powerful Combination.} 
Listing ~\ref{lstlisting: Example for the Combination of Annotations} shows the examples of the combinations of annotations to express the restrictions for the input parameter.
In particular, at the end of the example, we can see the power of an expression that further combines these annotations, \texttt{@Element, @SubClass, @Field, @Max, @Min}: 
this combination constrains the second element of the input array to be of type \texttt{Student} and whose field \textit{age} is  between 5 and 22.
\subsection{Implementation}
\label{subsection: Implementation}

\ToolName{} is built on the top of \texttt{Soot} and annotations are parsed using Java reflection technology.

\textbf{(1) Compatibility for existing annotations.}
The annotations we design reflect the meaning of basic expressive semantics.
The program under test can use these annotations.
At the same time, this is not required.
We collect common annotations and try to unify annotations that have the same semantics as the ones we have designed.
For compatibility, we have made efforts including  followings:
(1) ignore the package name and focus only on the class name; 
(2) ignore the case of the class name; 
(3) identify similar annotations, such as \textit{@Length} and \textit{@Size}; 
(4) parse the specific annotations of the annotation designed in this paper, such as annotations \textit{@Email}, \textit{@NotEmpty}, and \textit{@Digit}, which can essentially be represented by the annotation basis \textit{@Patter} with regular expressions expressing  specific meaning, which will be easily parsed compatibly.

\textbf{(2) Validation.}
As part of the code, annotations can be checked for its syntax by the compiler, but usually not for semantics.
Therefore, it is necessary to check for semantics. 
We will check whether these annotations are applicable to a variable, such as the annotation \textit{@Pattern} modifying an integer variable is an invalid message, thereby eliminating invalid constraints; 
annotations and their attributes will be ignored if there is a conflict between them, such as the value of the annotation \textit{@Min} modifying the same variable greater than the value of annotation \textit{@Max} or annotation \textit{@Min} modifying an array variable (for its length) but with a negative value.
Furthermore, as the numeric type contains types of different precision, an implicit message is that the upper and lower bounds of each numeric variable cannot exceed the range that the JVM can represent.

\textbf{(3) Test Data Generation.}
The annotation modifying a parameter of a method can be used to generate the input data of the test case for the method.
Our expectation is that test case parameters should match normal business values to improve their readability, etc. 
However, it does not mean that there is no need for values other than annotation constraints, because the test case needs to test not only the normal situation, but also the abnormal data situation to ensure its robustness. 
Therefore, when we generate test cases, we will not only include test input that satisfies the annotation constraint(s), but also other test input, such as the random value does not satisfy the constraint.

(i) For numeric types, we sort the values in the annotations (maximum or minimum, or multiple intervals), while the maximum and minimum values of each type are added as upper and lower bounds respectively. 
Each boundary is used as a basis for bucketing, and random values in each bucket are assigned to the parameters as representatives. 
(ii) The constraint for the String type variable is the regular expression.
And \texttt{RgxGen} ~\cite{RgxGen} from Github is used to generate string values based on regular expressions in this paper as its friendliness to complex ones ~\cite{JustinStr}. 
And a string candidate set are provided to complement random strings to express a wider range of string formats such as URL, XML, JSON, escape characters, etc.
(iii) And the annotation modifying a field can be used to construct the instance of the class it belongs to. 
If a field with an annotation could be assigned a setter method, then the method could be used to help construct the object. 

\textbf{(4) Assertion Generation.}
Although the above examples of annotations are modifying parameters, they can also modify the return variables smoothly,
So they can be used to generate assertions if they can pass the verification mentioned above.
The following code snippets show assertions for different return variables. 
\begin{lstlisting}
@Pattern("[0-9]{8,13}")
String foo1(...){String s; ... return s;}
// Assertion for foo1: 
{
    String s =  foo1(...);
    boolean b = Pattern.matches("[0-9]{8,13}",s);
    assert(b);
}

@Min(0)  @Max(22)
int foo2(...){int i; ... return i;}
// Assertion for foo2:
{ int r = foo2(...); assert(0 <= r && r <= 22); }
\end{lstlisting}
In addition, the annotation that modifies a field can also be used to generate assertions, particularly after a method call that includes an assignment statement for the field.

\section{Evaluation}
We will evaluate our approach from the following research questions.

\begin{itemize}
\item \textbf{RQ1:} How compatible are the annotations designed by us with existing annotations?
\item \textbf{RQ2:} Can existing approaches generate efficient test data to hit different spaces divided by variable constraints?
\item \textbf{RQ3:} Can \ToolName{} find defects in real word programs?

\end{itemize}

\subsection{Setup}
We  have randomly selected 5 projects\footnote{ I. https://github.com/geraldvaras/betterbanking;\\II. https://github.com/zanata/zanata-platform;\\III. https://github.com/prjrfco/Algamoney-Api;\\IV. https://github.com/fpoljcic/AuctionApp;\\V. https://github.com/Vokovalenko16/SpringREST;} from GitHub using annotations from the lib $javax.validation.constraints$ and built them ourselves.

\textbf{Setup for RQ1.}
In order to test the compatibility of our annotations with the existing annotations, we use the approach described in Section 3.4.1 for the selected Java programs in order to use the annotation information in the programs as much as possible for test case generation.
We then count all the annotations that modify fields, parameters and return values, and mark the annotation classes that we can use for test case generation and their usage in the programs.

\textbf{Setup for RQ2.}
To evaluate whether the input test data generated by \ToolName{} is more in line with the business requirements, we randomly select 15 numerical annotation usages on method parameters, which are mainly used to modify the model data, as a test set to observe the distribution of the values automatically generated by each tool (\texttt{EvoSuite} with default configuration and \texttt{Randoop} with $-attempted-limit=500$).

\textit{Combination Tool Selection.}
Although there are various test generation tools, they are basically based on two categories of methods, random and SBST. 
In this paper, we select the two most representative tools for comparison for the second research question, \texttt{Randoop} and \texttt{EvoSuite}.

\subsection{Compatibility}
\label{subsection: Compatibility}

\begin{figure}[t]
\centering
\includegraphics[scale=0.48]{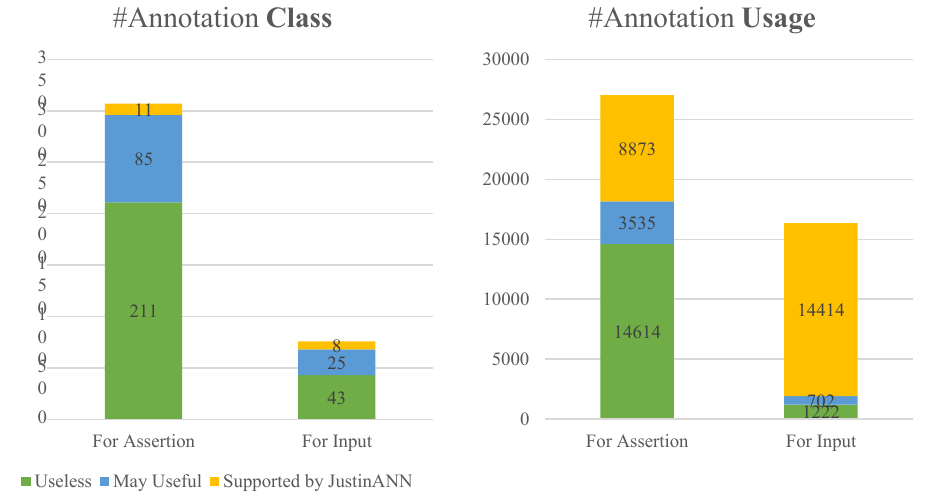}
\caption{Annotation Used in Programs}
\label{figure: Annotation Used in Programs}
\end{figure}

Figure ~\ref{figure: Annotation Used in Programs} shows the result of the annotations used in the programs.
The bar on the left represents the number of annotation classes, while the bar on the right represents the number of annotations used in the programs, including those for both return variables (For Assertions) and input parameters (For Input).
Among them, the ones which can not be used to guide test generation definitely are marked as \textit{Useless}, while the ones are marked as \textit{May Useful} which are not supported by \ToolName{} currently but may provide guidance.
It can be seen that a total of 96 annotations are used to modify return variables and fields, which can be used for the guidance of the test generation. 
\ToolName{} is compatible with 11 (11/(11+85))=11.5\%) of them, which have been used 8873 (8,873/(8,873+3535)=71.5\%) times.
In addition, for the annotations modifying parameters, which can be used for test data generation, \ToolName{}  is compatible with 8 (8/(8+25))=24.2\%) annotations of them, which have been used 14,414 (14,414/(14,414+702)=95.4\%) times.

\begin{figure*}[t]
\centering
\includegraphics[scale=0.75]{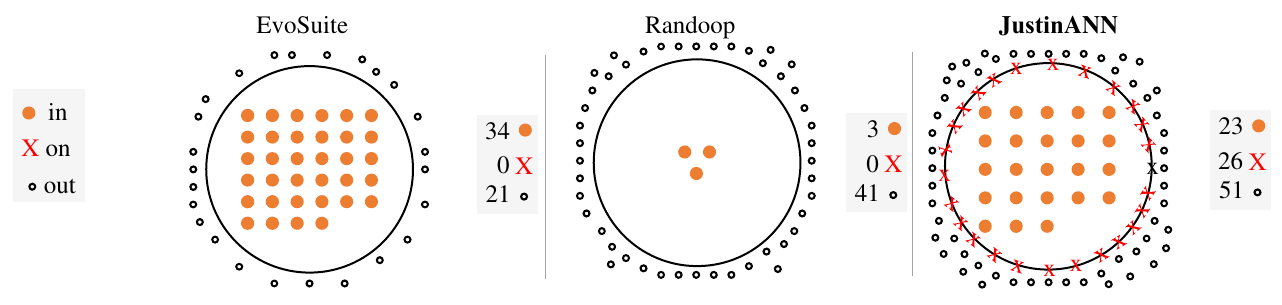}
\caption{Distribution of Test Data Generated by Different Tools. Legends,\texttt{in,on,out}, indicate that the generated test input is inside the given boundary, on the boundary, or outside the boundary.}
\label{figure: Distribution of Test Data Generated by Different Tools}
\end{figure*}

\begin{figure}[t]
\centering
\includegraphics[scale=0.7]{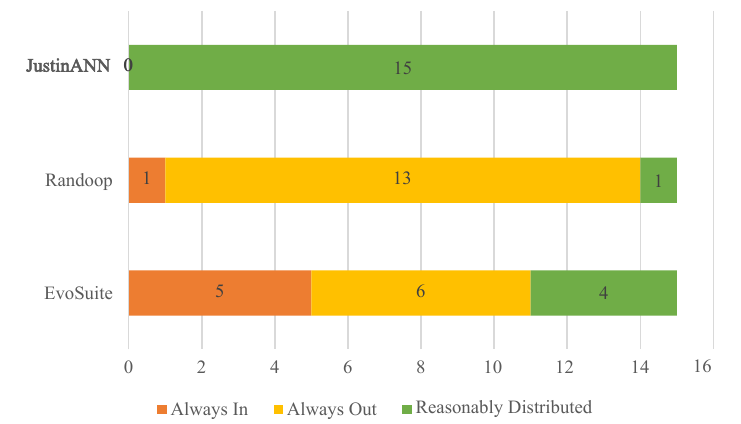}
\caption{Distribution of Input Data in Test Suites for Each Parameter Constrained by the Annotation. \texttt{Always In (Out)} means that, given a range for an input parameter from the annotation, the test input generated by the tool is always in (out of) the range. \texttt{Reasonably Distributed} means the inputs covers both values in and out of the range. }
\label{figure: Awayls}
\end{figure}

\begin{center}
\noindent\fbox{
    \parbox{.92\linewidth}{
\textbf{Answer to RQ1.}
\ToolName{} can be compatible with many annotations to guide the generation of test cases, especially those frequently used in the programs.
}
}
\end{center}

\subsection{Test Data Distribution}
As mentioned earlier, \ToolName{} has derived many annotations modifying parameters.
For the 15 numerical annotation usages, we manually count the distribution of test data generated by each tool.
The range of parameters restricted by annotations is used as the baseline. 
We divide the test data for these parameters into three categories, which are inside the boundary value, outside the boundary value and exactly on the boundary value, as shown in Figure~\ref{figure: Distribution of Test Data Generated by Different Tools}.
We can see that neither \texttt{EvoSuite} nor \texttt{Randoop} can cover the boundary values and \texttt{Randoop} has only 3 (3/44=6.8\%) of the 44 test cases in the range, while \ToolName{} can have a much better boundary coverage for them.
\texttt{Randoop} has difficulty generating function parameters within bounds because it uses a random strategy to generate input values as explained before.

Furthermore, although \texttt{EvoSuite} seems to cover both inside and outside the boundary test inputs, we have found that for a given parameter constrained by the annotation, the test data generated by it is almost (((5+6)/15 = 60\%)  either inside or all outside the boundary. 
It is shown in Figure ~\ref{figure: Awayls}.
The situation in \texttt{Randoop} is even worse (((1+13)/15 = 93.3\%).
Meanwhile, the test inputs generated by \ToolName{} are always cover both the values in and out of the constraint, which is more reasonable for testing of both normal and exceptional paths.

\begin{center}
\noindent\fbox{
    \parbox{.92\linewidth}{
\textbf{Answer to RQ2.}
Compared to existing tools, given a parameter constrained by the annotation, the test input generated by \ToolName{} is most evenly distributed in, out and on the boundary of the range; i.e., for a given requirement, it always generates test data in, on and out of the bounds, which is reasonable and effectively matches the intent of unit testing.
}
}
\end{center}

\subsection{Bug Finding}
We use \ToolName{} to test popular open source programs and find 26 real defects confirmed by developers, where 6 are fixed already as shown in Table ~\ref{table:Defects}.
In addition to open source programs, \ToolName{} is also used in industrial practice, discovering thousands of real defects.

\begin{table*}
    \caption{Confirmed Defects Found by \ToolName{} in Open Source Community. (Commit, Merge or Pull in the ID Column: Fixed)}
    \label{table:Defects}
    \begin{center}
    \begin{tabular}{c|c|c|c|c|c}
        \hline
        \textbf{Project} & \textbf{Version} & \textbf{Description} & \textbf{\#GitHub Star} & \textbf{\#Defect} &
        \textbf{ID(s)} \\
        \hline
            commons-cli\cite{Apache-Commons-CLI} & 1.0 & \makecell[c]{Provides a simple API \\ for  a Command Line Interface}  & 330 & 1 & commit(*55886e)\\
            \hline
            groovy\cite{groovy}  & 2.5.6 & \makecell[c]{A powerful multifaceted programming \\ language for the JVM platform } & 16.5k & 1 & pull(1643)\\
            \hline
            libgdx \cite{libgdx}  & 1.10.0 &  Java game development framework  & 22.8k & 1 & 6709\\
            \hline
            hutool \cite{hutool}  & 5.7.15 &  A set of tools that keep Java sweet & 28.5k & 4 & 1975,1982,1980,1981\\
            \hline
            asm \cite{asm}  & 5.1 & \makecell[c]{Java bytecode manipulation \\ and analysis framework}  & 16.5k & 1 & {mergerequests(330)}\\
            \hline
            openjdk\cite{openjdk}   & \makecell[c]{8u292\\7u75\\12.0.2} &   Open source Java Development Kit (JDK) & 18.5k & 17 & \makecell[l]{commit(*12bd18, *a404a9, *ba7d11), 8278186,\\8278993,8279129,8279128,8279198,8279218,\\8279336, 8279341,8279342,8279422,8279423,\\8279424,8279362}\\
            \hline
            bishengjdk\cite{bishengjdk}  & 1.8 & OpenJDK customized version  & 502 & 1 & I4MWI1\\
             \hline
              \hline
            Total &&&&26 & \\
        \hline
    \end{tabular}
    \end{center}
\end{table*}

\textbf{Case Study.}
The annotation is helpful in finding bugs for program analysis or verification tools \cite{FindBugs,FindBugs_Annotations,errorprone,SPF}.
It is also useful for finding bugs through test case generation.
Listing~\ref{lstlisting:Case Study String} shows an example from the JDK.
The buggy method \textit{parseIdFromSameDocumentURI(...)} tries to parse a \texttt{String} parameter \textit{uri} with a specific format to obtain an ID.
The method deals with three cases: \textbf{(I)} empty string (line 3), \textbf{(II)} string containing at least one character (line 6) and \textbf{(III)} string whose substring starts with ``xpointer(id("" after its character is discarded (line 7).
It will throw an exception if the branch condition in line 7 returns a true value (case \textbf{(III)}) but the parameter \textit{uri} does not contain a character `'' (apostrophe symbol), as the variable \textit{i2} can be assigned  with the value -1 (line 9), causing a fatal exception (\texttt{StringIndexOutOfBoundsException}) after the \textit{subString(...)} invocation in the next line.

If there is an annotation \textit{@Pattern} list to express the different cases as shown in the example (line 15), then we can get three regular expressions to guide the test case for the method.
The test data corresponding to the three cases may be shown in lines 17-19.
They can quickly cover all statements and branches when taken as the input data of the test cases, thus triggering string index out-of-bounds exception.
It is not always easy to analyse all cases using static analysis, especially for the most complex case (case \textbf{III}), which is the only case where the bug is triggered.

\begin{center}
\noindent\fbox{
    \parbox{.92\linewidth}{
\textbf{Answer to RQ3.}
\ToolName{} has the ability to detect real defects ( 26 in popular open source programs).
}
}
\end{center}

\begin{figure}[t]
    \centering

\begin{lstlisting}[
caption={Bug Finding Example with Annotations},
label={lstlisting:Case Study String}]
// issue (ID: 8278186) in openjdk
public static String parseIdFromSameDocumentURI(String uri) {
    if (uri.length() == 0) {
        return null;
    }
    String id = uri.substring(1);
    if (id != null && id.startsWith("xpointer(id(")) {
        int i1 = id.indexOf('\'');
        int i2 = id.indexOf('\'', i1+1);
        id = id.substring(i1+1, i2); // may be -1, StringIndexOutOfBoundsException
    }
    return id;
}
//With @Pattern Annotations
@Pattern("^$") @Pattern("^[\s\S]xpointer(id($") @Pattern("^[\s\S]+$") String uri

//Test Data 1 : uri = "";
//Test Data 2 : uri = "lxpointer(id(xdf";
//Test Data 3 : uri = "vaifv09r43r";
\end{lstlisting} 
\end{figure}

\section{Threats to Validity}
In evaluation, randomness can lead to difficulty in reproducing experimental results.
In addition, the approach of generating test cases based on annotations relies on the quality of the annotations, and annotations themselves are difficult to be tested effectively.
In other words, it is difficult for us to test whether an annotation is really correct, especially if it has no relationship to other annotations, causing that the generated test inputs and assertions are not always as expected.
Moreover, the goal of our approach is to generate real input and assertions that can cover the requirements of business programs, but this does not mean that it can effectively improve test coverage (such as statement coverage, branch coverage).
At present, annotations in this paper can not support the expression of dependencies between multiple parameters, such as two parameters representing an array variable and its length variable respectively.

\section{Related Work}
In this section, we will discuss related work for Java program from  perspectives involved in this paper, test data and assertion in automated test case generation.
\subsection{Test Data Generation}
\textbf{(1) Random approach}.
Random is perhaps one of the simplest forms of test data generation.
For example, \texttt{Randoop} is a feedback-directed random testing tool that generates unit test cases for Java programs ~\cite{Randoop}.
In fact, other tools or techniques are also inseparable from the random approach, such as search-based software testing (SBST) approach also requires random values to fill uninstructed inputs or directly assign to parameters in constraints that are time-consuming to solve.
\textbf{(2) SBST}. 
\texttt{Evosuite} is originally based on genetic algorithms (GA) to generate test cases, and later inherits
the dynamic symbolic execution approach to effectively improve coverage \cite{EvoSuite_dse}, which is also used in \texttt{UnitTestBot} \cite{utbot} with program analysis.
\textbf{(3) Mutation}. 
\texttt{PIT}, \texttt{Major}, and \texttt{Javalanche} are some of the most popular mutation testing tools, which aim to measure the effectiveness of the test cases by changing random parts of the software under test ~\cite{PIT, Major,Javalanche}, known as mutations, and the technique has some limitations, such as its seed quality, computational cost, time consumption and etc.
There are also works focusing on the test data generation for the specific input, such as String parameter \cite{JustinStr}, array parameter \cite{data_generation_array}, object field \cite{data_generation_evoObj},  and hybrid ones \cite{data_generation_Hybrid}.
It is difficult for them to consider the readability, which will affect the maintainability of test cases.
Studies also report their limitation in the understandability or readability ~\cite{ReadabilityStudy1,ReadabilityStudy2}.
Our approach leverages the existing annotation information in the program to guide the generation of test cases, which is a better solution to this problem.

\subsection{Assertion Generation}
The recent study highlights the importance of high-quality, complex assert statements when detecting real faults \cite{Assertion_study}.
Traditional testing tools generate assertions with the results obtained after the test cases are executed first or relying on heuristics or ``intelligent" randomness \cite{Randoop,EvoSuite,utbot,agitar}. 
With the development of deep learning (DL) in recent years, it is utilized increasingly to improve software quality.
\texttt{ATLAS} is the first DL-based assertion generation approach  which applies Neural Machine Translation (NMT) to aim at predicting a meaningful assertion for the given method \cite{ATLAS}.
Information Retrieval (IR) techniques have been widely applied in different software quality tasks \cite{retrieval1,retrieval2, retrieval3}, and also used for assertion generation \cite{Assertion_DeepLeaning}, There is also a research work  automatically improving assertion oracles with genetic evolutionary approach \cite{GAssert}.
Compared with them, our approach utilizes the commonly used annotations in programs containing abundant domain specifications, especially Spring-based service ones, to generate test cases that are more in line with business requirements, and more easily for programmers to understand and fewer consumption to maintain.

\section{Conclusion}
This paper presents \ToolName{}, with an annotation-guided approach to generating meaningful test cases for Java programs, which is compatible with the annotations commonly used in programs with rich domain specification.
Annotations that modify parameters, fields and return variables can guide the generation of test data and assertions.
We design 7 annotations and 4 combination rules to express the properties for both primitive and complex type variables.
Experiments show that many existing annotations are compatible with our approach, which makes it easier to generate test data in, out and on the requirement domain, and that it also helps to find program defects.

In the future, we will consider combination theory to guide test generation for methods with multiple parameters modified by annotations, and consider assertion generation for methods that do not return a variable.
And we will combine our approach with existing tools to generate more domain-sensitive test cases.

\newpage

\bibliographystyle{unsrt}
\small
\bibliography{8Ref}


\end{document}